\documentclass[preprint,showpacs,preprintnumbers,amsmath,amssymb]{revtex4}

\usepackage{graphicx}
\usepackage{dcolumn}
\usepackage{bm}

\begin{document}

\preprint{APS/123-QED}

\title{Observation of Macroscopic Structural Fluctuations in bcc Solid $^4$He}

\author{O. Pelleg, M. Shay, S. G. Lipson, and E. Polturak}
\email{emilp@physics.technion.ac.il}
\affiliation{ Physics Department, Technion, Israel Institute of
Technology, Haifa 32000, Israel}

\author{J. Bossy}
\affiliation{CNRS, BP 166, 38042 Grenoble Cedex 9,
 France}
\author{J.C. Marmeggi}
\affiliation{CNRS, BP 166, 38042 Grenoble Cedex 9,
 France}
\author{H. Kentaro}
\affiliation{CNRS, BP 166, 38042 Grenoble Cedex 9,
 France}
\author{E. Farhi}
\affiliation{Institut Laue-Langevin,  BP 156, 38042 Grenoble Cedex
9,France}
\author{A. Stunault}
\affiliation{Institut Laue-Langevin, BP 156, 38042 Grenoble Cedex
9,France}


\date{\today}

\begin{abstract}
We report neutron diffraction studies of low density bcc and hcp
solid $^4$He. In the bcc phase, we observed a continuous dynamical
behaviour involving macroscopic structural changes of the solid.
The dynamical behaviour takes place in a cell full of solid, and
therefore represents a solid-solid transformation. The structural
changes are consistent with a gradual rotation of macroscopic
grains separated by low angle grain boundaries. We suggest that
these changes are triggered by random momentary vibrations of the
experimental system. An analysis of Laue diffraction patterns
indicates that in some cases these structural changes, once
initiated by a momentary impulse, seem to proceed at a constant
rate over times approaching an hour. The energy associated with
these macroscopic changes appears to be on the order of kT. Under
similar conditions (temperature and pressure), these effects were
absent in the hcp phase.
\end{abstract}
\maketitle
\section{Introduction}
It is well known that quantum solids exhibit unique properties
concerning internal motion of atoms, properties which are most
prominent in solid Helium at the lowest density. Examples include
the rapid atomic exchange in $^3$He\cite{guyer}, extremely fast
annealing of bcc $^4$He\cite{sanders}, and mass diffusion at a
rate comparable to that of a liquid\cite{berent}. Recently,
experimental reports have appeared suggesting that at temperatures
below 0.2K, superfluidity may occur in hcp phase of solid $^4$He
\cite{goodkind,Chan&kim1,Chan&kim2}. In the following, we show
evidence of yet another unusual behaviour of bcc solid $^4$He,
namely that inside a crystal, macroscopic structural changes can
occur which may continue unabated for a long time. Our proposed
interpretation of these observations is that the structural
changes are gradual orientation changes of macroscopic crystalline
grains, with the energy involved in the process being comparable
to that of thermal fluctuations. A comparison of experiments which
we carried out over the last several years suggests that these
structural changes are triggered by sporadic momentary vibration
in the laboratory, some of which couple to the cell containing the
solid. In section II, we describe the experimental results. The
interpretation of these results is presented in section III.
\section{Experimental}
\begin{figure}[h!]
\includegraphics[width=9cm]{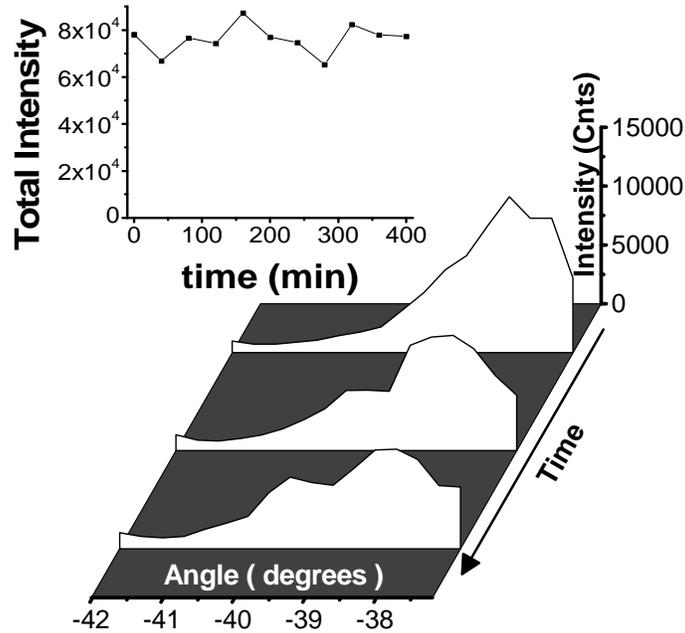}
\caption{\label{fig1} Time dependence of elastic scattering
spectra of a large bcc crystal, around the (110) reflection.
Successive spectra were obtained with the triple axis
spectrometer, an hour apart. The inset show the total integrated
intensity vs. time.}
\end{figure}

In this work, we investigated structural changes of low density He
crystals using neutron diffraction. For the most part, the results
discussed here refer to two separate experiments performed at the
Institut Laue Langevin. We first discovered these effects during
an experiment on the IN-14 triple axis spectrometer, using a beam
of cold neutrons having $\lambda$ = 4.05~$\rm \AA$ and a cross
sectional area of several cm$^2$. In this experiment, we grew He
crystals inside a spherical cell having a volume of 8.9 cm$^3$.
The spherical shape was chosen to maintain the same scattering
intensity regardless of its orientation relative to the beam.
Crystals of bcc $^4$He were grown at a constant temperature,
typically 1.650 K. Once the cell was completely filled with solid,
the temperature of the cell was further reduced by 10 mK, to
ensure that all the He in the filling line also solidified and the
crystal in the cell is well isolated. Subsequent experiments were
done on the solid phase under conditions of constant density and
temperature ($\pm$ 1mK). During the crystal growth, we monitored
the intensity of the (110) Bragg reflection. At this stage, the
diffraction pattern consisted of a single large intensity peak,
which indicated that the cell contained one large single bcc
crystal. Soon after starting the procedure to orient the crystal
in the beam by rotating and tilting the cryostat, we found that
the crystal broke up into several large, slightly misoriented
pieces. Further diffraction scans revealed a very interesting
dynamic effect; the various sub-crystals changed their orientation
and relative size continuously, in a seemingly random fashion.
Typical elastic diffraction spectra near the (110) Bragg
reflection, taken an hour apart, are shown in Figure 1. One can
see that with time, spectral weight is shifted away from the main
peak, indicating that parts of the crystal change their spatial
orientation. The inset in Figure ~\ref{fig1} shows the total
integrated intensity of these spectra, which remains almost
constant as a function of time. Hence, these spectra reflect
spontaneous changes involving the whole crystal, with each
sub-peak in the diffraction spectrum representing the orientation
of a \emph{macroscopic} grain, some cm$^3$ in size. We emphasize
that the whole process takes place in a cell which is completely
filled with solid. This solid-solid transformation process
persisted unabated for the duration of the experiment, close to a
week.

\begin{figure}[h!]
\includegraphics[width=8cm]{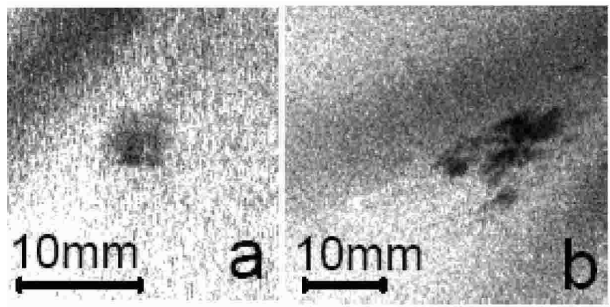}
\caption{\label{fig2} (a) Bragg spot of a small bcc single
crystal.(b) The same spot after thermal cycling. }
\end{figure}

In order to learn more about this effect, we performed a second
experiment, using a white beam neutron Laue Camera (S42). The
white beam has neutrons with 1$\rm~\AA~<~\lambda~<~3.5 \AA$ and a
small divergence(8'). This small divergence allows for the
resolution between crystalline grains misaligned by less than
1$^\circ $. In order to learn how to interpret the Laue images, we
first grew a small single crystal in a separate cell with a 0.2
cm$^3$ volume. Figure~\ref{fig2}(a) shows a Laue image with a
Bragg spot of a bcc single crystal. One can see that the intensity
within the Bragg spot is quite uniform. In our experience with
observing He crystals in our optical cryostat\cite{tuvy2}, cycling
the crystal through the hcp-bcc transition results in a break up
of the single crystal into smaller grains. After carrying out this
procedure, it is seen in Figure~\ref{fig2}(b) that the Bragg spot
is no longer uniform, showing streaks. These streaks are due to
Bragg reflections from crystal grains which are now slightly
misoriented. The larger size of the spot reflects the larger
rocking curve of the whole crystal.

\begin{figure}[h!]
\includegraphics[width=9cm]{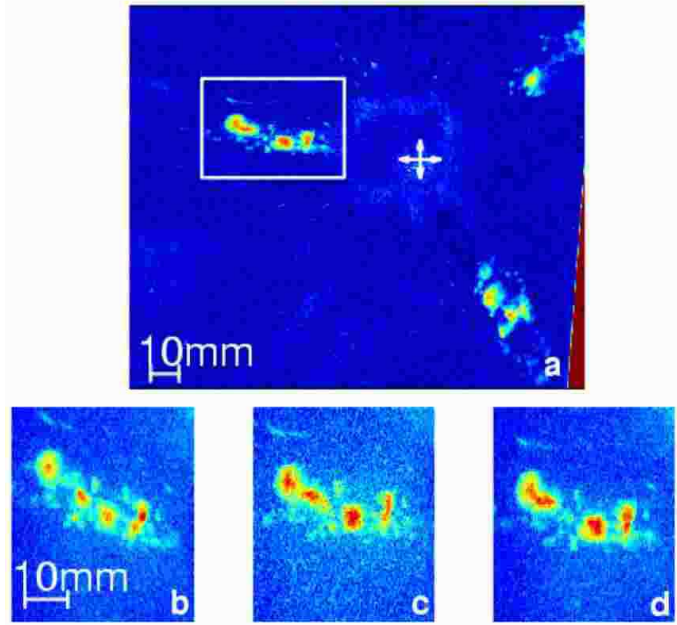}
\caption{\label{fig3}(color online) Top panel: Laue image of a
large bcc crystal, showing 3 large (110) spots. The cross marks
the center of the beam. Lower panels: sequential (enlarged) images
of one of the Bragg spots (boxed in the top panel)taken 15 min.
apart.}
\end{figure}

In the second part of the Laue camera experiment we grew another
bcc solid, this time in the large spherical cell. The solid was
grown intentionally to produce several slightly misoriented
crystals. The Laue diffraction pattern is shown in
Figure~\ref{fig3}. Background scattering due to the cell walls was
substracted from the image. In contrast to X-rays, neutrons
penetrate through the whole sample, and images are acquired in
transmission mode. With the large cell, the crystal size along the
beam direction is larger than the cross section of the neutron
beam. Since the neutrons penetrate the whole crystal, one expects
the Laue diffraction spots to become extended along the radial
direction relative to the center of the image. In this case, the
size of the diffraction spot on the image, $w$, is linear related
to the actual size of the sample. More accurately, for a perfect
single crystal of diameter $D$, the size of the Bragg spot along
the radial direction of the image is $w= D\tan(2\theta)+ h$, where
$h$ is the width of the beam and $\theta$ is the angle satisfying
the Bragg condition. The angular width of the Bragg spots is
proportional both to the rocking curve of the crystal and the
width of the beam. In Fig.~\ref{fig3}(a), three large spots can be
seen, roughly 120$^\circ $ apart, with some sub-structure inside
each one. By simulating the diffraction process\cite{filhol}, we
identified these 3 large spots as (110) Bragg reflections, with
the crystal oriented so that its [111] direction is at an angle of
18$^\circ $ to the neutron beam. Due to the Debye-Waller factor of
solid He, (110) Bragg reflections are much more intense than any
higher order reflection and therefore most likely to be seen. For
a real crystal with a finite rocking curve, the Bragg spot will be
larger than $w$, which characterizes a perfect single crystal.
However, the sum of the individual sizes of the small features
inside each large Bragg spot should equal $w$. We checked and
found that this indeed is the case. According to what we learned
from the images shown in Fig. 2, the sub structure inside each
large spot in Fig.~\ref{fig3}(a) indicates that the crystal is
made of several slightly misoriented grains. The dynamic behavior
of the crystal can then be followed by monitoring the changes of
such images as a function of time. Figures~\ref{fig3}(b),
~\ref{fig3}(c), and~\ref{fig3}(d) show three successive images of
one of the spots, taken 15 minutes apart. 15 minutes were
sufficient to obtain an image with a good contrast, with typical
features having an intensity of 5-10 times the noise level. It is
evident that some features inside the large spot change with time,
either losing intensity at the expense of others which become more
intense, or changing their position from one frame to the next.
Similar type changes occur simultaneously in all three large
spots. Intensity variations associated with these changes are on
the order of $10{\%}$ of the total intensity of the large spot.
This implies that the size of the crystal grains which undergo
these changes is a fraction of 1 cm$^3$ (in contrast to the triple
axis experiment, the beam cross sectional area of the Laue camera
is smaller, so we see only part of the spherical cell). As in Fig.
1, these changes indicate that some crystal grains inside the cell
change their orientation with time.

Structural changes of an imperfect crystal can be a natural part
of an annealing process. At a given temperature, annealing
continues for as long as thermal activation can induce further
changes in the crystal. The temporal extent of this process was
investigated by several authors\cite{sanders, beamish, hiki,
berent}. At temperatures above 1K, which are relevant to our
experiment, it was found that the typical time scale to anneal a
crystal is less than one hour. In the narrow range in which the
bcc phase exists, the most efficient way to anneal the crystal is
to heat it up until one just reaches the melting curve. Under
melting conditions, the crystal is surrounded by superfluid which
can relieve stress very efficiently by transporting heat and
matter. Stress relief by this method was demonstrated in several
experiments in which the crystal was subjected to mechanical
deformation\cite{sanders, berent, berent2}. In order to check
whether annealing is the process seen here, we raised the
temperature by 10 mK, and held the crystal on the melting curve
overnight, a very long time by the the criteria described above.
Both during and after the annealing, the crystal continued to show
random structural changes at apparently an undiminished rate for
the duration of the experiment (close to two weeks). We conclude
that these \textbf{macroscopic} structural changes are not a part
of an annealing process.

Next, we try to identify the source of the applied stress which
induces the structural changes. There are two possible sources of
stress, of thermal and mechanical origin. Thermally induced stress
results from the relatively large thermal expansion of solid He.
Since crystals are confined by rigid boundaries, any temperature
change will generate stress. Thermally induced stress was observed
by Hiki and Tsuruoka\cite{hiki}, who followed the dynamics of
dislocations in He crystals using ultrasonic techniques after
intentional abrupt temperature changes of 50 mK. A similar study
was carried out by Beamish\cite{beamish}, who observed the
relaxation of stress induced by large ultrasonic pulses upon
heating the crystal by several hundred mK. In all our experiments
at the ILL, both long term and short term temperature stability
was within $\pm$ 1mK. Therefore, thermally induced stress is a
priori much smaller. The temperature profile within the cryostat
in both experiments discussed here (IN-14 and the Laue experiment)
was identical, with the temperature gradient in the cryostat tail
static to within the same precision over the duration of the
experiment. Yet, the size of of the structural changes seen in
these two experiments is very different. Finally, we may add that
in our previous experiments, no structural changes were observed.
In those experiments\cite{tuvy}, the cell was better vibration
isolated, while its thermal stability was the same. We can
therefore conclude that thermally induced stress is
inconsequential in the present context.

Let us now consider mechanically generated stress. First, there is
no static applied stress in our experiment. However, a fluctuating
mechanical stress does exist due to external vibrations coupled to
the solid cell. The importance of this source of stress can be
appreciated by comparing the structural changes observed during
the three different experiments which we carried out at the ILL.
In the first experiment\cite{tuvy}, the cell was rigidly attached
to a 1/2" diameter stainless steel tube. In this experiment,
despite the fact that the cryostat was continously rotating on the
spectrometer, we saw no structural changes. Next, in the Laue
camera experiment, the spherical cell was clamped in a double
plate structure attached to 4 thin wall 1/8" tubes, which are much
less rigid than the 1/2" tube. The cryostat holding the cell was
static. In this experiment, we saw structural changes amounting to
orientation changes of 2$^{\circ}$-3$^{\circ}$ over the duration
of the experiment. Finally, in the experiment done on the IN-14
triple axis spectrometer, the double plate support structure
attaching the spherical cell to the 4 thin wall 1/8" tubes was
removed in order to reduce the scattering background. The cell was
therefore more susceptible to vibrations. Furthermore, during
position changes, the cryostat together with the spectrometer were
subject to a higher acceleration than used previously\cite{tuvy}.
In this experiment, we observed cumulative orientation changes of
10$^{\circ}$-20$^{\circ}$. The importance of the acceleration was
tested directly by monitoring the Bragg reflections from the
crystal while changing the acceleration of the cryostat (and of
the experimental cell).  When the motion of the cryostat was
"soft", meaning very low acceleration, we observed that the Bragg
 reflections of the crystal did not change with time, meaning
 that there were no structural changes. When the acceleration was
 increased, we immediately observed that the crystal broke up
 exactly in the way shown in Fig. 1. Since the acceleration
 is of short duration, it follows that large enough momentary
 mechanical acceleration leads to crystalline changes
 (all of this was done under isothermal
 conditions).The comparison of all these experiments strongly
suggests that the continuous solid-solid transformation process
reported here is driven by mechanical vibrations of the cell.

The cell containing the solid He is a spherical metallic shell,
attached to the end of a thin long rod (1m) suspended from a top
plate of the cryostat. If the cryostat is suddenly moved, it will
excite pendulum-like vibrations of the cell. The pendulum
frequency comes out about 0.5Hz, which is within the band of
typical building vibrations discussed below\cite{baklakov}.
Another possibility is an excitation of transverse elastic
oscillations of the rod which couple into the cell. We estimate
the resonant frequency of the lowest mode around 100Hz.  The metal
cell itself is not affected by these vibrations, however the solid
He, which is compressible, will be affected by the stress exerted
on it by the cell wall. If the vibration is strong enough, the
stress will exceed the critical shear stress and structural
changes will follow.

Typical low frequency vibrations spectrum in a laboratory has a
1/f$^\alpha$ dependence, with 2 $> \alpha >$1 \cite{baklakov}. To
affect the solid, the vibration amplitude of the cell must exceed
the critical shear stress of the solid (3$\times$10$^{4}$
dyne/cm$^2$\cite{sanders}). Because of the 1/f dependence of the
vibration spectrum, vibrations having an amplitude large enough to
shear the solid will be less frequent. Focusing on the Laue
experiment, we now estimate the likely interval between successive
vibrations which induce structural changes. If the exposure time
of a Laue image is long compared with the interval between
vibrations, individual diffraction spots such as the small
features shown in Figs.~\ref{fig3}(b)-(d) will be able to shift
back and forth on the image under a sequence of uncorrelated
vibrations which cause random changes during the exposure period.
The Laue images exhibit the intensity integrated over the exposure
time. If the exposure time is long enough compared with the
interval between structural changes, the image should look
uniform, like in Fig.~\ref{fig2}(a). If however the exposure time
is short in comparison, the images should show isolated
diffraction spots. We found that after an exposure time of 10
hours, the image indeed looks quite uniform, namely many
vibrations have occurred. Shorter exposure times of 15 minutes to
1 hour produce images such as Figs.~\ref{fig3}(b)-(d), showing
distinct diffraction spots, with some apparent changes from one
frame to the next. From these observations, we deduce that the
interval between vibrations in the Laue experiment is at least
several minutes.

\begin{figure}[h!]
\includegraphics[width=9cm]{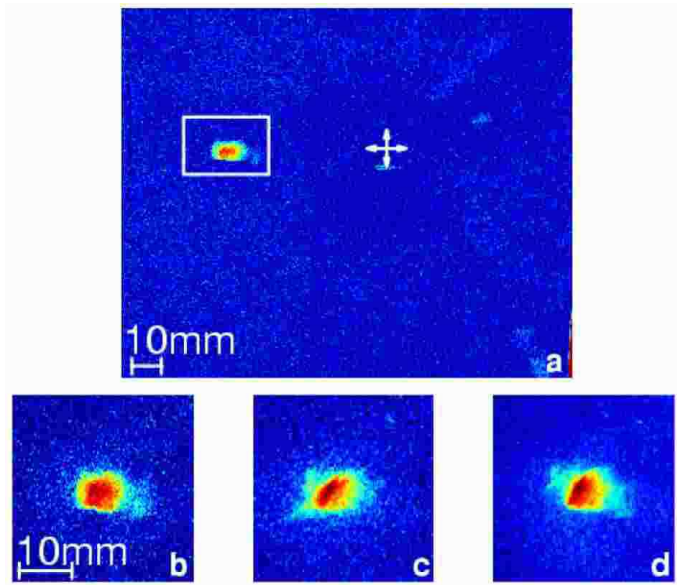}
\caption{\label{fig4}(color online) Top panel: Laue image of a
large hcp crystal, showing a single Bragg spot. The cross marks
the center of the beam. Lower panels: sequential images of the
Bragg spot taken 15 min. apart.}
\end{figure}

It is interesting to check whether this phenomenon occurs only in
the bcc phase or also in the hcp phase. To that end, we repeated
the experiment with a hcp helium solid grown at a constant
temperature of 1.80 K, which is still in the regime of low density
solid. In Figure ~\ref{fig5}(a) one can see a single Bragg
spot,which is uniform and \emph{small} compared to $w$. The fact
that only one Bragg spot was seen may indicate that the
polycrystal is made up of grains which are strongly misoriented,
so that only one grain satisfies the Bragg condition within the
spherical angle spanned by the image plate, while the other grains
do not and are therefore not visible. A similar image was obtained
with another hcp crystal, grown in the smaller cell. In contrast
to the bcc solid, the hcp Laue image changed much less with time
(see Figure~\ref{fig4}(b)-~\ref{fig4}(d)). These observations are
consistent with the work of Iwasa et al.\cite{iwasa2}, who studied
hcp $^4$He crystals using high resolution X-ray topography.
Similarly to our observations, these authors have not reported any
dynamic behaviour of the hcp solid.

\section{Discussion}

In this section, we outline one example of a classical mechanism
which is consistent with the structural changes observed in the
experiment. Analysis of the diffraction patterns from the Laue
camera experiment indicates that the bcc crystal is made of grains
separated by low angle (3$^{\circ}$-7$^{\circ}$) grain boundaries.
The shift of the diffraction spots with time as shown in
Fig.~\ref{fig3}(b)-~\ref{fig3}(d); indicates that the orientation
of some grains changes. Orientation changes could result either
through straightforward rotation of grains or from a gradual
translation of grain boundaries\cite{comment}. Brute force
rotation of an arbitrarily shaped solid grain within a solid
matrix involves massive displacements of atoms at the boundaries
through vacancy diffusion\cite{berent2,dyumin} and dislocation
climb. It requires a very specially arranged stress field.  In the
absence of such field, vacancies and dislocations will not flow in
a correlated way which is needed for a macroscopic rotation of a
grain. Consequently, rotation of grains is less likely than the
mechanism discussed below.

\begin{figure}[h!]
\includegraphics[width=8cm]{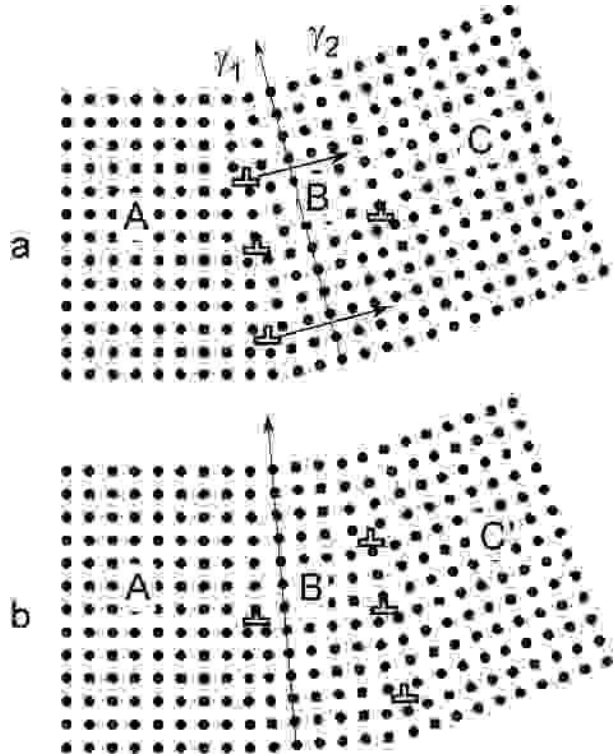}
\caption{\label{fig5}  Schematic illustration of the mechanism of
effective rotation of a crystal grain affected by progressive
motion of dislocations from grain boundary $\gamma_1$ to
$\gamma_2$. In (a), the orientation of grain B of the crystal is
close to that of grain C, while in (b), the orientation has
changed to that of A.}
\end{figure}

A mechanism which appears more plausible involves a gradual
translation of a low angle grain boundary inside the crystal, and
can in principle proceed via dislocation glide in a slip plane.
The Laue images are consistent with the solid being made up of
grains separated by simple tilt grain boundaries. These consist of
parallel edge dislocations of the same sign, spaced by $\delta y
\equiv b/\alpha$, with $\bf{b}$ the Burgers vector and $\alpha$
the misorientation angle. Figure~\ref{fig5} illustrates one
possibility how glide of individual dislocations can gradually
change the orientation of a grain. In the figure, dislocations
move from grain boundary $\gamma_1$ to grain boundary $\gamma_2$
at some rate. As a result, the misorientation angle of $\gamma_1$
decreases and that of $\gamma_2$ increases. Consequently, section
B of the crystal, bounded between grain boundary $\gamma_1$ and
$\gamma_2$, changes its orientation continuously starting from
that close to grain C and ending with that of grain A. In order
for this process to occur, dislocations must be able to detach
themselves from the grain boundary $\gamma_1$. A perfect tilt
grain boundary is stabilized mainly through the elastic force
between  edge dislocations resulting from their mutual stress
field. It is however possible for the dislocation to wander away
from the grain boundary through nucleation of kink pairs
overcoming the Peierls barrier, which advance the dislocation an
atomic distance at a time in the slip plane. To examine this
possibility, we evaluate the probability of formation of kink
pairs. A classical expression estimating $V_k$, the formation
energy of a pair of kinks\cite{Seeger,hiki} is
\begin{equation}\label{2}
  V_k =  W_k + 2E_Pl_c
\end{equation}
Here $W_k=(4a_0/\pi)(E_PE_0)^{1/2}$ is the energy of a kink,
$E_P=(a_0 b/2\pi)\tau_P$ is the height of the Peierls potential
barrier, with $E_0=2\mu b^2/\pi p$ the self energy per unit length
of a dislocation, $\mu$ is the shear modulus and $\tau_P$ the
Peierls stress. The critical distance separating the two kinks so
that they do not annihilate, $l_c$, is given by $l_c = a_0/2 \pi
\times (E_0/E_P)^{1/2} ln(32 E_P/a_0 b \tau_c)$. For bcc solid
$^4$He, the shear modulus $\mu$= 2.4$\times$10$^8$
dynes/cm$^2$\cite{greywall}, the lattice parameter
$a_0=$4.11$\bf\AA$, and the Peierls stress $\tau_P$ is 10$^{-4}$ -
10$^{-5} \mu $, depending on the particular type of
dislocation\cite{hiki, iwasa}. For edge dislocations in bcc solid
$^4$He, the Burgers vector is $b=\frac{\sqrt{3}}{2}a_0$. Taking an
average value of $\tau_P =3\times$ 10$^{-4}\mu $ ,we find
$V_k$=0.29K. Quantum corrections\cite{iwasa}, namely zero point
motion, enhance the exchange probability of atoms and decrease
$V_k$ further. At the temperature of our experiment, T=1.65K,
there should be a sizable population of thermally excited kink
pairs and dislocation glide should be possible. This conclusion is
supported by experimental findings; at temperatures similar to
ours, Sanders et al.\cite{sanders} could not detect any increase
in ultrasonic attenuation in bcc crystals under dynamic
deformation. This implies that dislocations generated by a moving
piston were able to reach the walls and annihilate at the same
rate they were created. Similar conclusion regarding the large
mobility of edge dislocations was arrived at by Hiki and
Tsuruoka\cite{hiki}.Torsional oscillator studies by Miura et
al.\cite{miura} show that in pure crystals below 1K, pinning of
dislocations occurs at the intersection with other dislocations
which are immobile. In an impurity free crystal above 1K, where
all the dislocations are highly mobile, there are practically no
pinning sites and dislocation should be able to wander from one
grain boundary to another.

\begin{figure}[h!]
\includegraphics[width=9cm]{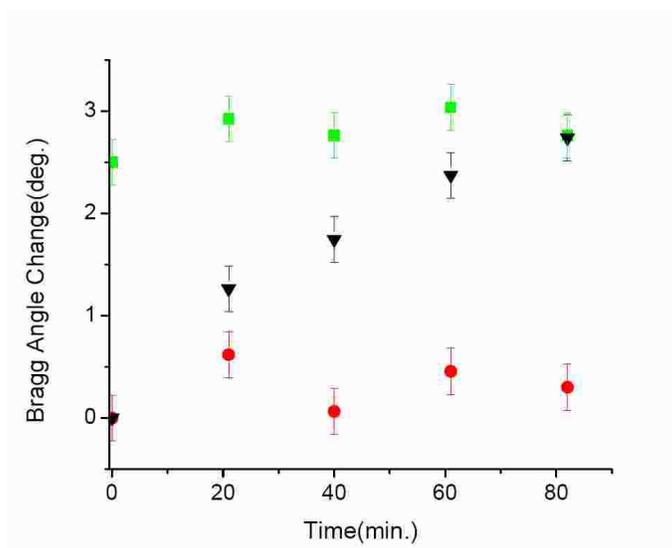}
\caption{\label{fig6} (color online) Measured orientation changes
of a crystal grain, expressed through the change of the angular
position of its Bragg spot vs. time. The rotating grain
(triangles) splits off a static grain (circles) and changes its
orientation until reaching the orientation of another static grain
(squares). The error bars represent the uncertainty of reading off
the position of the grain on the image plate.}
\end{figure}

Crystalline orientation changes consistent with the mechanism
shown in Figure~\ref{fig4} and discussed in the preceding
paragraph were actually observed in our experiment. In some cases,
we were able to resolve a steady progress of a diffraction spot on
the image plate from one frame to the next over several
consecutive frames. We show these data in Fig. 6. The moving
diffraction spot splits off from a static spot, and terminates in
another static spot. The intensity of the moving spot indicates
that the size of the crystal grain associated with it is about
0.07 cm$^3$. The total time span over which this steady progress
was observed was over one hour.  It seems that in this particular
case, the interval between some momentary vibrations was long
enough so that we could observe the whole relaxation process.
Referring to Figure~\ref{fig4}, the rate at which dislocations
cross from one grain boundary to the other, $\dot{n}$, is given by
$\dot{n}=(L/b)\dot{\alpha}$, with $L$ the size of the grain
boundary. For the data shown in Fig. 6, we calculate
$\dot{n}\approx$ 10 dislocations/sec. This is a very small number
compared with a typical dislocation density in He crystals of
10$^4$-10$^5$ cm$^{-2}$\cite{iwasa2}. Given that the energy
involved in the motion is that of the kinks, $V_k$=0.29K, the
implication of our observation is that macroscopic structural
changes can take place with an energy input of order kT. Hence,
the term "macroscopic structural fluctuations" seems appropriate
to describe the behaviour of bcc crystals. In closing this
paragraph, we stress that the mechanism illustrated above is
proposed on grounds of plausibility and is not a full theoretical
model, so that other interpretations of these unusual effects are
possible.

Finally, we address the different behaviour of the bcc and hcp
crystals. This difference can perhaps be understood in the
following way: In the bcc structure, the edge dislocation with the
lowest energy has a Burgers vector $\textbf{b}=\frac{1}{2}$[111].
The slip planes for this type of dislocation are \{110\}, \{112\},
and \{123\}, offering 12 different slip planes for any particular
[111] dislocation. Hence, easy motion of low angle grain
boundaries can take place easily in practically any direction. In
contrast, in the hcp structure there is only one preferred slip
plane, \{0001\}, which implies that motion of grain boundaries in
hcp will be strongly restricted relative to the bcc phase.

\section{Acknowledments}
We are grateful to Frederic Thomas, Shmuel
Hoida, M. Blein, and P. Thomas for their invaluable help, and N.
Gov for discussions. This work was supported in part by the Israel
Science Foundation and the Technion VP Research fund.



\begin{thebibliography}{99}
\label{Bib}
\bibitem{guyer} R. A. Guyer, R. C. Richardson, and L. I. Zane, \textit{Rev.
Mod. Phys.} \textbf{43}, 532 (1971).
\bibitem{sanders} D. J. Sanders, H. Kwun, A. Hikata and C. Elbaum,
\textit{Jour. Low Temp. Phys.} {\bf 35}, 221 (1979), \textit{Phys.
Rev. Lett.} {\bf 40}, 458 (1978).
\bibitem{berent} I. Berent and E. Polturak , \textit{Phys. Rev. Lett.}
\textbf{81}, 846 (1998).
\bibitem{goodkind} J. M. Goodkind,\textit{Phys. Rev. Lett.} {\bf 89}, 095301(2002).
\bibitem{Chan&kim1} E. Kim and M.H.W Chan, \textit{Nature} \textbf{427}, 225(2004).
\bibitem{Chan&kim2} E. Kim and M.H.W. Chan, \textit{Science} \textbf{305},1941 (2004).
\bibitem{tuvy2}T. Markovich and E. Polturak, \textit{Jour. Low Temp. Phys.}
\textbf{123},  53 ( 2001).
\bibitem{filhol} Simulations were done using "Orient Express", a computer program written
by A. Filhol.
\bibitem{beamish} J. R. Beamish and J. P. Franck, \textit{Phys. Rev.} {\bf B28}, 1419(1983).
\bibitem{hiki} Y. Hiki and F. Tsuruoka, \textit{Phys. Rev.} {\bf B89}, 696(1983).
\bibitem{berent2} I. Berent and E. Polturak, \textit{Jour. Low. Temp. Phys.} {\bf 112}, 337(1998).
\bibitem{tuvy} T. Markovich, E. Polturak, J.Bossy and E. Farhi, \textit{Phys. Rev. Lett.}
\textbf{88}, 195301 (2002).
\bibitem{baklakov} B. Baklakov, T. Bolshakov, A. Chupyra, A. Erokhin, P. Lebedev, V. Parkhomchuk, Sh. Singatulin, J. Lach and V. Shiltsev \textit{Phys. Rev. Special Topics - Accelerators and Beams} {\bf 1},031001(1998).
\bibitem{iwasa2} I. Iwasa, H. Suzuki, T. Suzuki, T. Nakajima, I.
Yonenaga, H. Suzuki, H. Koizumi, Y. Nishio, and J. Ota,
\textit{Jour. Low. Temp. Phys.} {\bf 100}, 147(1995).
\bibitem{comment} For completeness, there is a possibility for the long time
dynamics being due to the solid behaving as a glass. However, some
key characteristics of glass-like behavior are missing in our
observations. First, the diffraction pattern is characteristic of
a single crystal or an aggregate of a few crystals, each one of a
very high quality. The diffraction pattern does not show any
amorphous characteristics of glass like materials. Second,
thermally activated relaxation in glasses leads to a gradual
transition from an amorphous to crystalline state. This relaxation
proceeds only in one direction, and simply stops if the
temperature is reduced. Application of mechanical stress leads to
amorphization, and again, the process goes only in one direction.
Neither of these relaxation types fits our observations, in which
we see crystal transform back and forth around the same state.

\bibitem{dyumin} N. V. Dyumin, N. V. Zuev, S. V. Svatko, and V. N. Grigoriev,
\textit{ Sov. Jour. of Low Temp. Phys.} \textbf{17}, 458 (1991).
\bibitem{iwasa} I. Iwasa, N. Saito, and H. Suzuki, \textit{Jour. Phys. Soc. Jap.} {\bf 52}, 952(1983).
\bibitem{Seeger} A. Seeger, Phil. Mag. {\bf 1}, 651 (1956).
\bibitem{greywall} D. S. Greywall,\textit{Phys. Rev.} {\bf B13}, 1056(1976).
\bibitem{miura} Y. Miura, K. Ogawa, K. Mori, and T. Mamiya, \textit{Jour. Low. Temp. Phys.} {\bf 121}, 689(2000).

\end{thebibliography}

\end{document}